\documentclass[aps,prb,twocolumn,showpacs,preprintnumbers,amsmath,floatfix]{revtex4}
\usepackage{graphicx}
\usepackage{dcolumn}
\usepackage{subfigure}
\usepackage{wrapfig}
\usepackage{cancel}
\usepackage{color}
\usepackage{bm}

\begin{document}

\def\k{{\bf k}}
\def\rr{{\bf r}}
\def\q{{\bf q}}
\newcommand{\blue}{\textcolor{blue}}
\newcommand{\red}{\textcolor{red}}
\newcommand{\green}{\textcolor{green}}

\title{ \bf  Theory of thermal conductivity in extended-$s$ state superconductors: application to ferropnictides}
\author{V. Mishra$^1$, A. Vorontsov$^2$, P.J. Hirschfeld$^1$, and I. Vekhter$^3$}
\affiliation{$^1$Department of Physics, University of Florida,
Gainesville, FL 32611, USA\\$^2$ Department of Physics, Montana
State University, Bozeman, MT 59717 USA \\$^3$ Department of
Physics and Astronomy, Louisiana State University, Baton Rouge, LA
70803, USA}
\date{\today}

\begin{abstract}
Within a  two-band model for the recently discovered ferropnictide
materials, we calculate the thermal conductivity assuming general
superconducting states of $A_{1g}$ (``$s$-wave") symmetry,
considering both currently popular isotropic ``sign-changing" $s$
states and states with strong anisotropy, including those which
manifest nodes or deep minima of the order parameter.  We consider
both intra- and interband disorder scattering effects, and  show
that in situations where a low-temperature linear-$T$ exists in
the thermal conductivity, it is not always ``universal" as in
$d$-wave superconductors.  We discuss the conditions under which
such a term can disappear, as well as how  it can  be induced  by
a magnetic field.  We compare our results to several recent
experiments.
\end{abstract}

\maketitle

\section{ Introduction}

\label{sec:intro}

The symmetry class of the newly discovered ferropnictide
superconductors\cite{ref:kamihara} is still controversial, due in
part to differing results on superfluid
density\cite{ref:Hashimoto,ref:Malone_Martin,ref:Hashimoto2,ref:Gordon,ref:Gordon2,ref:Fletcher},
angle-resolved photoemission
(ARPES)\cite{ref:Zhao,ref:Ding,ref:Kondo,ref:Evtushinsky,ref:Nakayama,ref:Hasan},
nuclear magnetic resonance
(NMR)\cite{ref:RKlingeler,ref:Zheng,ref:Grafe,ref:Ahilan,ref:Nakai,MYashima:2009},
Andreev
spectroscopy\cite{ref:Shan,ref:Chien,ref:Daghero,ref:Gonnelli},
and other experimental probes.  In some cases, these experiments
have been interpreted as implying the absence of low-energy
excitations, i.e. a fully developed spectral gap. In others,
low-energy excitations have been observed, and taken as indication
of the existence of order parameter nodes.  It may be that these
differences depend on the stoichiometry or doping of the
materials, which affects the pairing interaction, or sample
quality, or both.

As in other classes of potentially unconventional superconductors,
one's ability to identify the symmetry class of a candidate
material by observation of low-$T$ power laws in temperature
reflecting low energy quasiparticle excitations is limited by how
low in $T$ one can measure.  At intermediate temperatures,
variations of thermodynamic and transport properties can be
affected by details of band structure, elastic and inelastic
scattering, as well as the presence of thermal  phonons. Only at
the very lowest $T$ can one -- in principle -- extract direct
information on the order parameter structure. Thermal conductivity
measurements have played an important role in past discussions of
unconventional superconductivity
{\cite{YMatsuda:2006,Shakeripour:09}}, in part because they can be
extended to $T$ of order tens of mK.  In addition, such
measurements are distinguished because they are bulk probes, and
because they are unusually sensitive to the presence of order
parameter nodes. {If lines of nodes are present, the thermal
conductivity $\kappa(T)$ manifests a low-$T$ linear, in
temperature, term which is purely electronic in origin and is
associated with residual quasiparticle states at the Fermi level,
induced by disorder or a magnetic field.} If, in addition, the
order parameter averages to zero over the Fermi surface (as in the
$d$-wave case appropriate for the cuprates), this linear-$T$ term
in zero field is known to be ``universal", in the sense that its
magnitude is only weakly disorder dependent.

Very recently, several low-temperature measurements of thermal
transport have been made on the BaFe$_2$As$_2$ (Ba-122) material
doped with K\cite{Tailleferkappa,Ongkappa},
Co\cite{TanatarCo:2009} and Ni\cite{Likappa}, as well as on the
stoichiometric superconductor LaFePO\cite{Matsudakappa}. In the
case of the Ba-122 samples, either zero or very small linear-$T$
terms have been reported in zero field, leading to the conclusion
that there is a fully developed spectral gap in these
{materials\cite{note}} in these experimental works.  This is in
contrast to reports of power law temperature dependence in the
superfluid density measured on the same
materials\cite{ref:Hashimoto,ref:Malone_Martin,ref:Hashimoto2,ref:Gordon,ref:Gordon2},
as well as other strong indications of low-energy excitations. One
way to reconcile these experiments is to note that thermal
conductivity at mK temperatures probes lower energy scales than
those measured in other experiments to date; thus it is possible
that a band of low-energy excitations extends to very low
energies, but not all the way to the Fermi level, either due to an
intrinsic highly anisotropic order parameter with deep minima, or
a band of impurity states which lies at low but nonzero excitation
energies. Such impurity states can be produced, e.g., in
isotropic, sign-changing $s$-wave pair
state\cite{ref:Muzikar,ref:Mazindisorder,ref:Kontanidisorder,ref:YBang}
allowable in multiband systems if special conditions on the ratio
of intra- to interband scattering are met.  {However, a further
strong constraint from the thermal conductivity measurements is
that a significant linear-$T$ term in the thermal conductivity is
observed with the application of a small magnetic field of order
one Tesla and hence much below the upper critical field, $H_{c2}$.
This residual term grows with increasing field.} This would be
consistent with the existence of { quasiparticle states at low,
but finite energy}. Refs.
{\onlinecite{Tailleferkappa,TanatarCo:2009}} on K- and Co-doped
Ba-122 samples claimed that this enhancement is significantly
larger than that to be expected in the case of a conventional
$s$-wave superconductor. {In contrast, Ref. \onlinecite{Likappa}
came to opposite conclusions on the measurements of a Ni-doped
sample, and Ref.~\onlinecite{Tamegaikappa} reported a small but
significant linear-$T$ term in zero field in  Co-doped Ba-122.}

{A recent measurement of $\kappa(T)$ on the ferrophosphide
superconductor LaFePO finds a very large linear-$T$ term
\cite{Matsudakappa}. If this is interpreted as indicative of order
parameter nodes, it would be consistent with the linear-$T$
dependence in the superfluid density also observed for this system
\cite{ref:Fletcher}.}  Note that disorder in a sign-changing $s$
state cannot produce such a term in the superfluid density.  This
material is the only material yet discovered among the growing
ferropnictide family of superconductors whose undoped ``parent
compound" is superconducting at zero pressure. It is therefore
expected to be significantly cleaner than other superconductors
discussed here.  This may be relevant because it has been proposed
that disorder in highly anisotropic ``s-wave" (A$_{1g}$ symmetry)
states can ``lift" shallow nodes in the order parameter, leading
to a fully developed spectral gap\cite{Mishraetal:2009}. {The
authors of Ref. \onlinecite{Matsudakappa} note that, despite a
very sharp resistive transition, the cannot completely exclude the
possibility that the linear-$T$ term is partly extrinsic; however,
even in that case the dominant dependence of the thermal
conductivity on the magnetic field should come from the
superconducting phase.}


{There is a {developing} consensus that the gap changes sign
between the electron and hole Fermi surface sheets}. From the
theoretical standpoint, states with nodes or deep minima appear to
be quite natural. Several microscopic {theories} of the spin
fluctuation mediated pairing interaction in the ferropnictide
materials have attempted to {calculate} the momentum dependence of
the order parameter associated with the leading superconducting
instability. Using a 5 Fe-orbital model, Kuroki {\it et
al.}~\cite{ref:Kuroki} performed an RPA calculation of the
 interaction to construct a
linearized gap equation, and determined that the leading pairing
instability had $s$-wave ($A_{1g}$) symmetry, with nodes on the
electron-like Fermi surface (``$\beta$ sheets"). Wang {\it et
al.}~\cite{ref:Wang}  studied the same pairing problem  within a
5-orbital framework using the functional renormalization group
approach, also finding that the leading pairing instability is in
the A$_{1g}$-wave channel,  and that the next leading channel had
$B_{1g}$ ($d_{x^2-y^2}$) symmetry.    For their interaction
parameters, they found   no nodes on the Fermi surface, but
nevertheless a significant variation of the magnitude of the gap.
Graser et al also performed a 5-orbital RPA
framework\cite{ref:graseretal}, using the DFT bandstructure of Cao
et al.\cite{ref:Cao} as a starting point. These results indicated
that the leading pairing channels were indeed of $s$ ($A_{1g}$)
and $d_{x^2-y^2}$ symmetry, and that one or the other could be the
leading eigenvalue, depending on details of interaction
parameters. More recently, several authors have investigated the
factors including intrasheet Coulomb interaction, nesting of
electron pockets, and orbital character of pairing which can
influence order parameter anisotropy within these
models\cite{Maieranisotropy,Chubukovanisotropy,Berneviganisotropy}.
Other approaches have also obtained $A_{1g}$ gaps which change
sign between the hole and electron Fermi surface sheets but remain
approximately isotropic on each
sheet\cite{ref:Mazin_exts,ref:Chubukov_exts}.

In this paper we calculate the expected thermal conductivity in
superconducting states potentially appropriate to the
ferropnictide superconductors.  We adopt for convenience a
phenomenological  2-band model, allowing  order parameters on two
Fermi surface sheets representing the hole- and electron- doped
sheets found in density functional theory calculations for these
materials. We consider the region outside of the doping range
where superconductivity may coexist with antiferromagnetism. Our
model for disorder consists of terms allowing for scattering
within (intraband) and between (interband) Fermi surface sheets,
of arbitrary strength. This allows us to control the width and
position of the impurity band in both nodal pairing states and
those with a fully developed spectral gap, which we examine with a
view towards determining the size and universality of the
linear-$T$ term in $\kappa$ at the lowest temperatures. After
examining the zero-field situation, we discuss the effect of an
applied field.  To this end we adopt the method of
Brandt-Pesch-Tewordt to obtain predictions for the {widest}
possible field range.  We illustrate the various possibilities of
superconducting state and disorder types which allow the results
observed thus far in experiments.

\section{Model}
\label{sec:model} We begin by assuming a metallic system with two
bands 1 and 2, characterized by densities of states $N_1$ and
$N_2$ at the Fermi level, and a  pair interaction which is a sum
of separable terms,
\begin{eqnarray} V(\k,\k') &=& V_1\Phi_1(\k)\Phi_1(\k') + V_2
\Phi_2(\k)\Phi_2(\k')\nonumber \\&&
+V_{12}[\Phi_1(\k)\Phi_2(\k')+\Phi_2(\k)\Phi_1(\k')],\label{pairpot}
\end{eqnarray}
where $\Phi_i$ is  function of $A_{1g}$ symmetry depending on
momentum restricted to band $i=1,2$.

For disorder we will assume an orbital-independent matrix element
which scatters quasiparticles either within a given band with
amplitude $U_{ii}$, $i=1,2$, or between bands with amplitude
$U_{12}$.  As discussed in the appendix, we sum all single-site
scattering processes of arbitrary strength to obtain a
disorder-averaged Nambu self energy ${\underline\Sigma} = n_{imp}
\,{\underline{T}}$, where $n_{imp}$ is the concentration of
impurities. {For simplicity, we assume $U_{11}=U_{22}\equiv U_d$,
with equal densities of states $N_i=N_0$ throughout the paper. In
our preliminary considerations we restrict ourselves to purely
intraband scattering, $U_{12}=0$. The disorder \blue{is}
characterized by two intraband scattering parameters on each
sheet: $\Gamma_i\equiv n_{imp}/(\pi N_i)$ and $c_i=1/(\pi
N_iU_{ii})$; For our simple initial case with 2 symmetric bands we
set $\Gamma_{i}=\Gamma$ and $c_i=c$, $i=1,2$. The initial neglect
of interband scattering may be understood in zeroth order by
noting that a screened Coulomb potential with screening length of
order a unit cell size will generically have larger small-{\bf q}
compared to large-{\bf q} scattering. {The real situation is
somewhat more complex since the same orbitals contribute to both
electron and hole Fermi surface sheets \cite{ref:graseretal}, and
therefore a substitutional impurity, such as Co, may be expected
to produce a significant interband scattering component as well.
Hence we relax this requirement and below also analyze the regime
$U_{12}\simeq U_{d}$, and, in particular, the case $U_{12}=U_d$,
as discussed in Ref.\onlinecite{ref:Kontanidisorder}.  Weak
interband scattering $U_{12}\ll U_{11},U_{22}$ does not
qualitatively change the results obtained in the limit
$U_{12}=0$.}

The full matrix Green's function in the presence of scattering in
the superconducting state is given  by a diagonal matrix in
band space, as discussed in the appendix,
\begin{equation}
{\underline G}({\bf k}, \omega) = {\tilde\omega\tau_0
+\tilde\epsilon_\k\tau_3+\tilde\Delta_{\bf k}\tau_1\over
\tilde\omega^2-\tilde\epsilon_\k^2-\tilde\Delta_\k^2},
\label{Greensfctn}
\end{equation}
{where $\k = \k_i\in S_i$ is restricted to Fermi surface sheet
$S_i$ with $i=1,2$,} and the renormalized quantities
$\tilde\omega\equiv \omega-\Sigma_0$, $\tilde\epsilon_\k\equiv
\epsilon_\k+\Sigma_3$, $\tilde \Delta_\k \equiv
\Delta_\k+\Sigma_1$ also depend on the band indices through $\k$.
The $\Sigma_\alpha$ are the components of the self-energy
proportional to the Pauli matrices
$\tau_\alpha$ in particle-hole (Nambu) space.}   

{Below we focus on two quantities. It is useful to start with the
analysis of the total density of quasiparticle states (DOS)}
\begin{eqnarray} N(\omega) &=& -{1\over 2\pi} {\rm Tr}~
{\rm Im} \sum_{\k} {\underline G}(\k,\omega)\nonumber\\
&=& -{1\over 2\pi} {\rm Tr}~ \sum_i \sum_{\k_i} {\underline
G}(\k_i,\omega),
\end{eqnarray}
where the second equality indicates the explicit integration over
distinct Fermi surface sheets with momenta $\k_i$.
We will be comparing results for the DOS with thermal conductivity
$\kappa$ calculated {using the standard
approach}\cite{AmbegaokarGriffin},
\begin{eqnarray}
\kappa &=&\sum_{i} \frac{N_{i} v_{Fi}^{2}}{8}
\int_{0}^{\infty} d\omega \frac{\omega^2}{T^2} {\rm
sech}^{2}(\frac{\omega}{2~T}) \nonumber \\ &&\times\left \langle
\frac{1}{{\rm
Re}\sqrt{\tilde{\Delta}_{i}^{2}-\tilde{\omega_i}^{2}}}
\left[1+\frac{|\tilde{\omega_i}|^2-|\tilde{\Delta_i}|^2}{|
\tilde{\Delta}_{i}^{2}-\tilde{\omega_i}^{2}|}\right]\right
\rangle_{\phi}\label{eq:kappa1}.
\end{eqnarray}
Here $<..>_\phi$ is average over the Fermi surface sheet,
$\sum_{i}$ denotes the sum over the bands, and $v_{Fi}$ is the
Fermi velocity on sheet $i$. We assume cylindrical Fermi surfaces,
so that ${\bf v}_{Fi}$ is isotropic.

\section{Isotropic A$_{1g}$ states}

We first discuss the thermal conductivity in isotropic $s$-wave or
A$_{1g}$ states.  If the sign of the order parameter $\Delta$ is
the same on both sheets, and no magnetic disorder is present, the
low-$T$ thermal conductivity will be similar to classic
calculations for conventional superconductors, and yield an
exponential low-$T$ dependence for the electronic part.  In the
case of a sign-changing $s$ (``$s_\pm$") state proposed by Mazin
et al.\cite{ref:Mazin_exts}, the situation is more interesting.
Here one assumes an isotropic $\Delta_i$ on each sheet $i$, but
assumes that ${\rm sgn}\Delta_1=-{\rm sgn} \Delta_2$. In terms of
Eq. \ref{pairpot},  {we choose the functions $\Phi_i(\k)=1$ for
$\k\in S_i$ and zero otherwise, and fix the sign of $V_{12}$ to be
opposite to that of $V_1,V_2$} so as to induce a sign-changing
order parameter between the two sheets. In the clean case, we
continue to expect an exponential or full gapped dependence to the
thermal conductivity. On the other hand, in such systems ordinary
disorder is pairbreaking if it includes a strong interband
component\cite{ref:Muzikar,ref:Mazindisorder,ref:Kontanidisorder,ref:YBang}.
For some situations a low-energy impurity band may indeed give
rise to a linear term in the thermal conductivity.

\begin{figure}
\begin{center}
\includegraphics[width= 0.8\columnwidth]{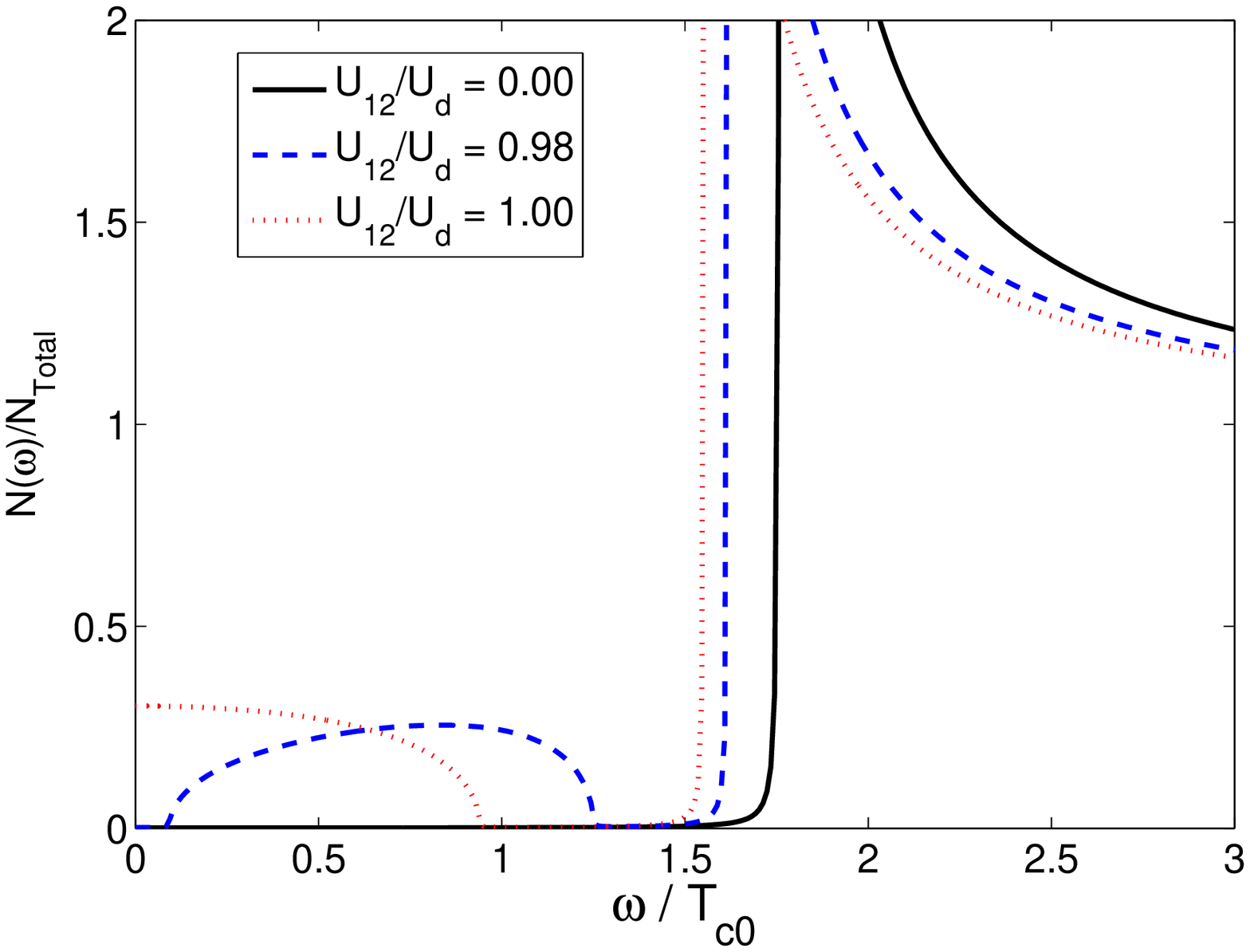}
\includegraphics[width= 0.8\columnwidth]{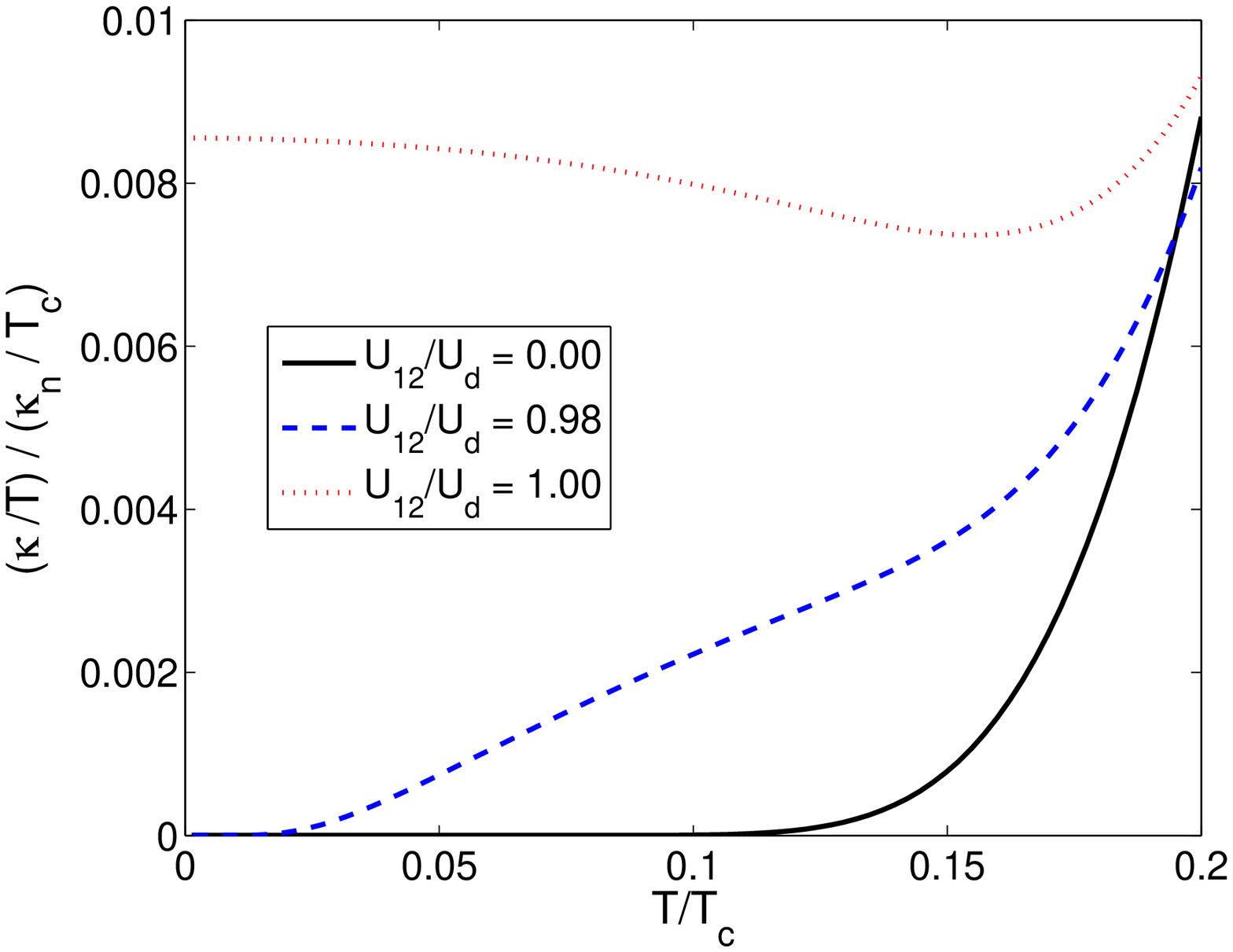}
\caption{Density of states (a) and  thermal conductivity (b) for
an isotropic $s_\pm$ state with $\Delta_1=-\Delta_2$, shown for
$U_{d} = U_{11}=U_{22}$ (intraband Scattering ) and scattering
rate parameters $c=0.07$ and  $\Gamma = 0.3 T_{c0}$ in cases (i)
weak intraband scattering only, $U_{12}/U_{d}$=0 (solid line);
(ii) pairbreaking scattering with midgap impurity band,
$U_{12}/U_d=0.98$ (dashed line); (iii) pairbreaking scattering
with impurity band overlapping Fermi level, $U_{12}/U_d=1.0$
(dotted line).} \label{fig:s_pm}
\end{center}
\end{figure}

For simplicity, we assume that $\Delta_2=-\Delta_1\equiv\Delta$,
and equal densities of states on the two bands. In Fig.
\ref{fig:s_pm}, we now illustrate the correspondence between the
formation of the impurity band in the fully gapped state, and the
creation of the linear term.  In the absence of interband
scattering, there is no pairbreaking in the sense of Anderson's
theorem, and the spectral gap in the DOS is identical to the
unrenormalized order parameter $\Delta$, corresponding to an
activated thermal conductivity $\sim\exp(-\Delta/T)$.  As
interband { scattering is increased, states are pulled down from
the continuum into the gap, creating eventually a band of midgap
states in the DOS, as shown. If there is still a narrow energy
range which is gapped near the Fermi level, this smaller gap
determines the slower but still exponential decrease of $\kappa$
at the lowest temperatures.  As soon as the impurity band of
midgap states overlaps the Fermi level, a linear term in $\kappa$
appears~\cite{ref:Mazindisorder,ref:Vorontsov}. {Note that a
significant interband component of scattering essentially equal to
the intraband component is absolutely necessary for this to occur,
which requires special conditions as described above}.

We note further that if one varies the concentration of impurities
in the situation with $U_{12}=U_d$, the change in the residual
density of states $N(0)$ is reflected directly in the slope of the
thermal conductivity, as shown in Fig. \ref{iso_universal}. This
is not surprising but is in dramatic contrast to the universal
(disorder-independent) behavior observed in $d$-wave
superconductors.  This raises the question of the degree of
universality of transport coefficients in pairing states which
have $s$-wave (A$_{1g}$) character but also display nodes.

\begin{figure}
\begin{center}
\includegraphics[width= 0.8\columnwidth]{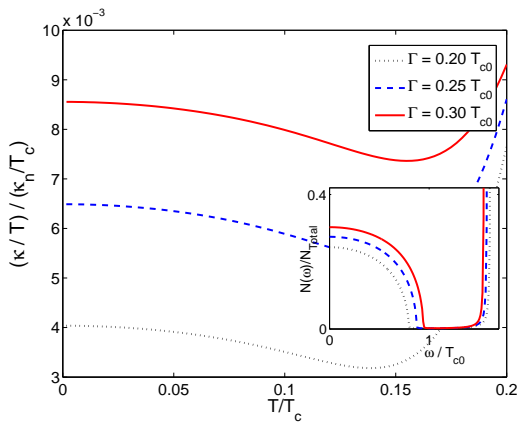}
\caption{Density of states (insert) and  thermal conductivity for
an isotropic $s_\pm$ state with  scattering parameters as Fig.
\ref{fig:s_pm} but $U_{12}/U_d$=1.0 and $\Gamma = 0.2,0.25,$ and
$0.3 T_{c0}$.}
\label{iso_universal}
\end{center}
\end{figure}

\section{ Anisotropic A$_{1g}$ states }  We now examine states
within the same A$_{1g}$ symmetry class, but where gap minima are
either very deep, with no sign change, or with actual sign change
(nodes).  To make contact with microscopic theory (see e.g. Ref.
\onlinecite{ref:graseretal}), we assume that one of the sheets (in
microscopic theory, the so-called ``$\alpha$" sheet around the
$\Gamma$ point) has an isotropic order parameter while the other
(the ``$\beta$" sheet around the M point) has a highly anisotropic
one.   In this case, the order parameters in the two bands are
given by
\begin{eqnarray}
\Delta_1 &=& - \Delta \\
\Delta_2 &=& \Delta_{iso} + \Delta_{ani} \cos 2\phi,
\label{eq:gps}
\end{eqnarray}
where $\phi$ is the angle around the electron Fermi surface sheet
2. {It useful to define the gap ratio in the electron-like band,
$r\equiv \Delta_{ani} / \Delta_{iso}$, so that $r>1$
($\Delta_{ani}>\Delta_{iso}$) gives a state with nodes in that
band, while $r<1$ ($\Delta_{ani}<\Delta_{iso}$) has none.}

{To obtain this order parameter from Eq. (\ref{pairpot}), we
choose}
\begin{eqnarray}
\Phi_1(\k)&=& \left\{ \begin{array}{cc} 1  &\k \in S_1\\0 &
\mbox{otherwise} \end{array} \right. \\ \Phi_2(\k) &=& \left\{
\begin{array}{cc} 1 + r_V \cos(2 \phi) & \k \in S_2\\0 &
\mbox{otherwise}
\end{array}\right.,
\end{eqnarray}
where {as before,} $S_1$ and $S_2$ represent the hole and electron
Fermi surfaces, respectively.  In the clean limit, the gap
equation is
\begin{eqnarray}
\Delta_{i} (\phi) &=& 2 \pi T \sum_{\omega_{n}} \Phi_i(\phi)
\nonumber
\\
&& \times\sum_{j} \int_{\phi'\in S_j} N_{j}V_{ij} \Phi_j(\phi')
\frac{\Delta_{j}(\phi^{\prime})}{\sqrt{\omega_{n}^{2}+\Delta_{j}^{2}(\phi^\prime)}}\label{eq:gp}
\end{eqnarray}
where $\omega_n$ are fermionic Matsubara frequencies. {Below we
adjust the value of $r_V$ to study a nodal system with $r=1.3$ and
an anisotropic state with no nodes with $r=0.9$.}


In the presence of disorder, we evaluate the impurity average self
energies $\Sigma_{i,\alpha}$ for both the bands, where again $i$
is the band index and $\alpha$ is the Nambu index. This
calculation is detailed in the Appendix, with the results
presented in Eqs.(\ref{Sigma10})-(\ref{SigmaD}). Since the
structure of the order parameter in the clean limit already
supports low-energy excitations, we first ignore the interband
scattering in the first analysis, and focus on the effects of
intraband scattering alone.   We define the conventional
renormalized quantities
\begin{eqnarray}
\tilde{\omega}_{i} &=& \omega - \Sigma_{i,0}~~~~i=1,2 \\
\tilde{\Delta}_i &=& \Delta_i + \Sigma_{i,1}\,, 
\label{eq:renorm}
\end{eqnarray}
where in each case the first subscript is a band index, while the
second one is a Nambu index.  Since the self energy is
\k-independent in this approximation, {we can associate the
self-energy with the renormalization of the isotropic component,
$\tilde{\Delta}_{iso}= \Delta_{iso} + \Sigma_{2,1}$, but this is
simply a matter of convenience.} The total thermal conductivity
comes from sum over both bands, but at very low temperatures the
contribution from the first band is very small due to the fully
developed gap assumed.
\begin{figure}
\begin{center}
\includegraphics[width= 0.83\columnwidth]{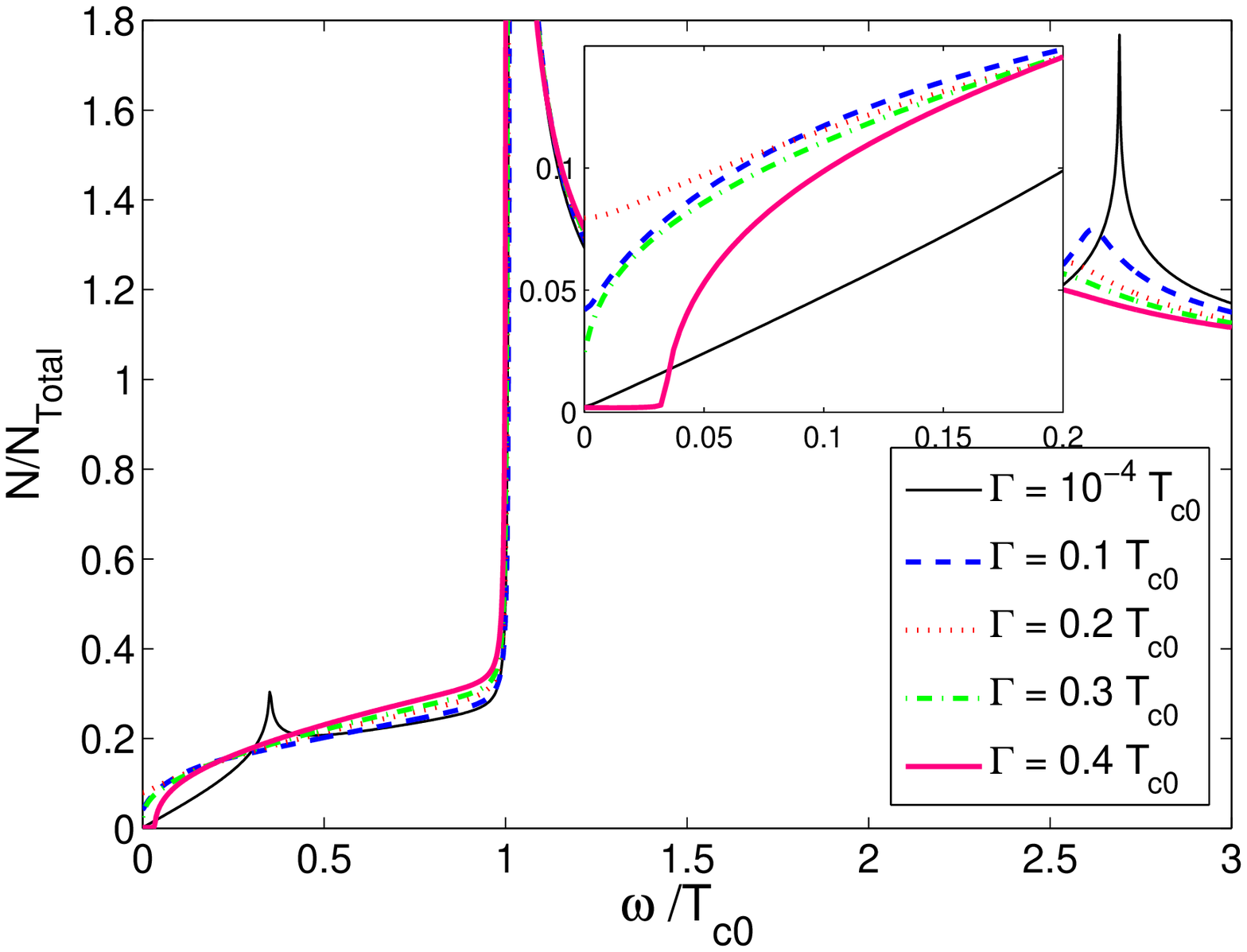}
\includegraphics[width= 0.9\columnwidth]{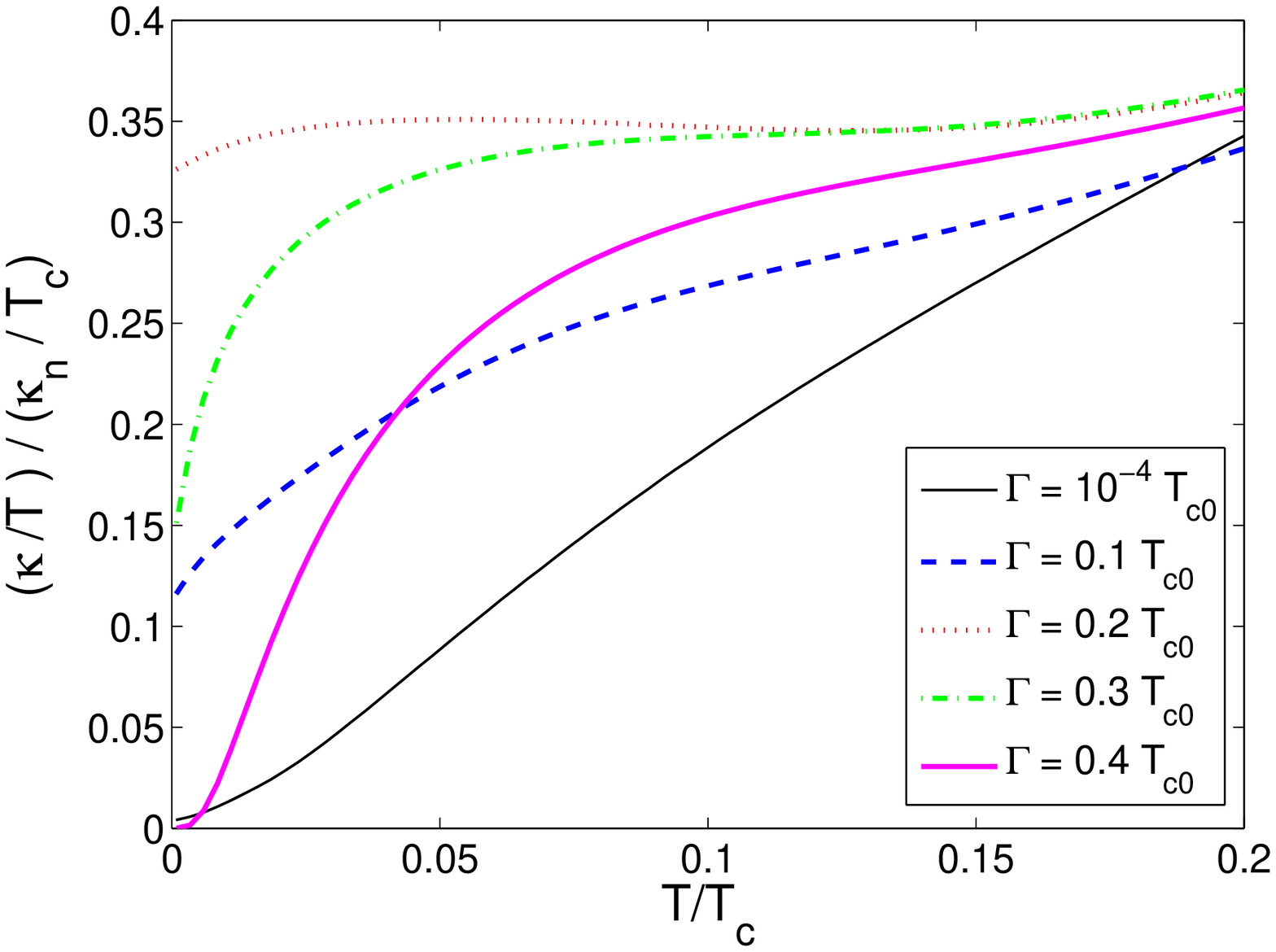}
\caption{Density of states $N(w)/N_{Total}$ (top) and normalized
thermal conductivity $(\kappa(T)/T)/(\kappa_n/T_{c})$ vs. $T/T_c$
{{for two band anisotropic model with isotropic order parameter on
sheet 1 and anisotropic order parameter $\Delta_1=-1.1~T_{c0}$ on sheet 1
and $\Delta_{iso}=1.3~T_{c0}$ , $r=1.3$ on sheet 2 (Eq. (\ref{eq:gps})).}} Results shown  for
various values of intraband scattering rate $\Gamma/T_{c0}$ and
$c=0.07$ , and no intraband scattering, $U_{12}=0$.}
\label{fig:kappa}
\end{center}
\end{figure}

\subsection{A$_{1g}$ states with nodes}

We first discuss the situation where $\Delta_2(\k)$ has {nodes but
a non-zero average over the Fermi surface}, and for concreteness
take $r=1.3$. The behavior of the low-$T$ thermal conductivity
with increasing disorder for a case with { individual scatterers
near the {strong potential} limit} is now shown in Fig.
\ref{fig:kappa}.

{The evolution of $\kappa(T)$  is very different from that for a
pure $d$-wave superconductor. This is clearly seen from the
evolution of the $T=0$ limit of the thermal conductivity. In the
pure case the linear term is nearly invisible. As intraband
disorder is increased, the linear term significantly increases in
magnitude, goes through a maximum and eventually disappears,
leading to an exponential temperature dependence.}  To some extent
this behavior can be understood by examining the corresponding
density of states, as shown in the upper panel; as disorder
increases, the nodal quasiparticle states are broadened and a
residual density of states appears,  but as disorder is increased
further the nodes are lifted and a fully developed spectral gap
appears, as discussed in Ref.~\onlinecite{Mishraetal:2009}.

{To clarify why there is no ``universal independence" of weak
disorder expected, e.g. for $d$ wave
superconductors\cite{PALee,Grafetal,PHNorman}, we plot in Fig.
\ref{fig:kappa_T0} }the value of the  asymptotic low-$T$ limiting
value of $\kappa/T$ as a function of disorder; there is, for this
case, no range of disorder where the behavior can in any sense be
called universal. {This result is somewhat similar to that in Ref.
\onlinecite{Maki_BM} where the effect of an orthorhombic
distortion on the in plane thermal conductivity of $YBCO$ was
studied. Note that in their case the $s-$wave component of the
$d+s$ order parameter breaks $A_{1g}$ symmetry in the ab-plane. In
our case $A_{1g}$ }is preserved because there are two $\beta$
sheets whose nodal structures are rotated with respect to each
other by 90 degrees \cite{ref:graseretal}.

\begin{figure}
\begin{center}
\includegraphics[width= 0.9\columnwidth]{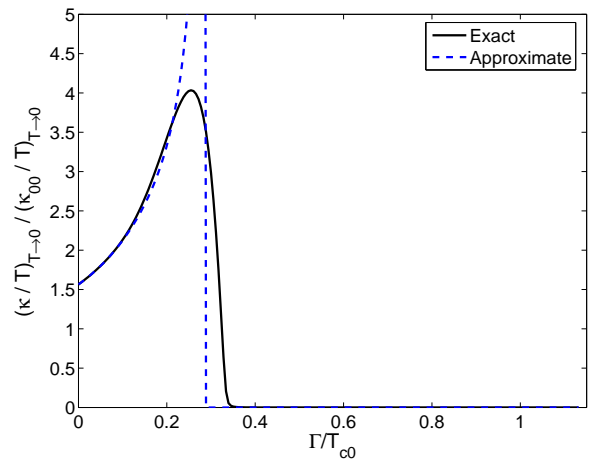}
\caption{Magnitude of linear-$T$ term in thermal conductivity in
limit $T\rightarrow 0$ for case with nodes $r=1.3$, plotted as a
function of intraband scattering rate $\Gamma/T_{c0}$ for
$c=0.07$. Solid line: exact numerical result. Dashed line:
analytical estimate from text, Eq. (\ref{eq:app2}).{{ Here
$\kappa_{00}$ is the thermal conductivity for a $d-$wave
superconductor, with order parameter
$\Delta(\phi)=\Delta_{ani}~\cos 2\phi$. $\Delta_{ani}$ is same as
anisotropic component in clean system for the model specified by
Eq. (\ref{eq:gps}).}}} \label{fig:kappa_T0}
\end{center}
\end{figure}

To analyze the origin of the breakdown of universality in the
anisotropic $A_{1g}$ state, {we evaluate the $T\rightarrow 0$
limit by replacing} the derivative of the Fermi function by a
delta function and {integrating over the energy} $\omega$. The
isotropic band 1 does not contribute to the thermal conductivity
at very low temperature because it is fully gapped. The main
contribution comes therefore from the nodal states from band 2.
{In contrast to} a $d$-wave superconductor, here the anomalous
self-energy $\Sigma_{2,1}$ is finite, therefore the compensation
between the density of states and the scattering rate does not
occur, and the universal behavior breaks down. As a result, the
position of the nodes on the Fermi surface shifts, and the slope
of the gap changes.}



{If we linearize the gap near the node,
 $\tilde{\Delta} (\phi) \approx k_F v_\Delta (\phi_0 - \phi)$,
where $\phi_0$ is the location of node, determined from
 $\cos 2\phi_0 = - [\Delta_{iso} +
 \Sigma_{2,1}(\omega=0)]/{\Delta_{ani}}$,
the renormalized gap slope is}
 \begin{eqnarray}
 v_\Delta &=& 2 k_F^{-1}~\Delta_{ani}~ \sin (2\phi_0 ) \\ &=& 2
 k_F^{-1}
 \sqrt{\Delta_{ani}^2-(\Delta_{iso}+\Sigma_{2,1}(\omega=0))^2}.
 \label{eq:gap_velocity}
 \end{eqnarray}
{Summing over the nodes, we find}
 \begin{eqnarray}
 \frac{\kappa}{T} \approx \frac{N v_{F}^{2}\pi}{3}
 \frac{2}{k_F v_\Delta},
\label{eq:app2}
 \end{eqnarray}
{which has precisely the same form as the well-known $d$-wave
result} (to leading order in $v_\Delta/v_F$), {\it except} that
the gap velocity $v_\Delta$, which is unrenormalized by disorder
in the $d$-wave case, is  strongly disorder dependent {here} due
to the nonzero off-diagonal impurity self-energy $\Sigma_{2,1}$ in
Eq. \ref{eq:gap_velocity}. {The increase in the residual thermal
conductivity is therefore due to the flattening of the gap in the
near nodal region before the system becomes fully gapped at higher
impurity scattering rates. The absence of the residual linear term
in this picture is only consistent with a {sufficiently} high
disorder, when the spectral gap is finite.}

\subsection{Anisotropic states with deep gap minima}  
\begin{figure}
\begin{center}
\includegraphics[width= 0.9\columnwidth]{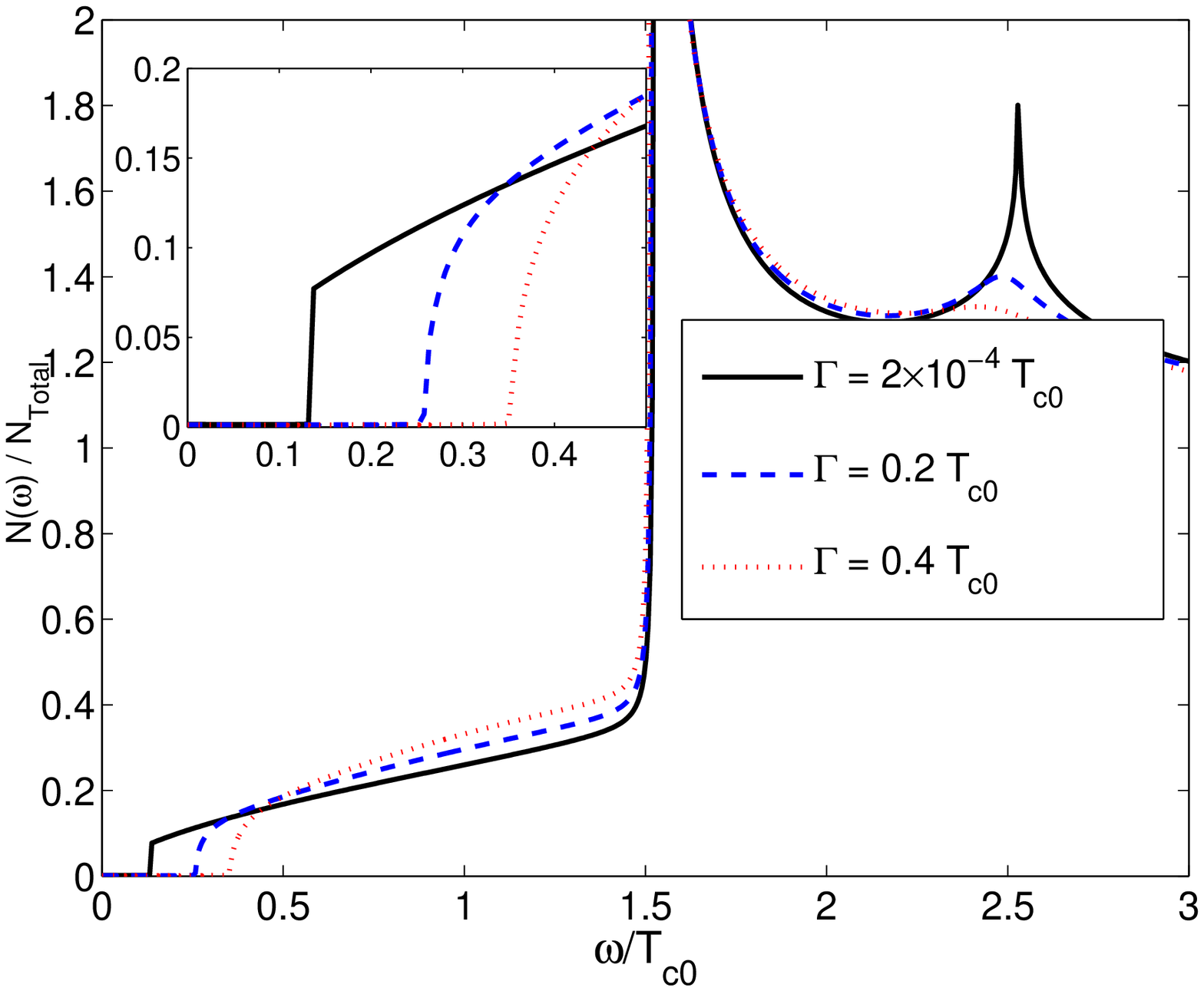}
\includegraphics[width= 0.9\columnwidth]{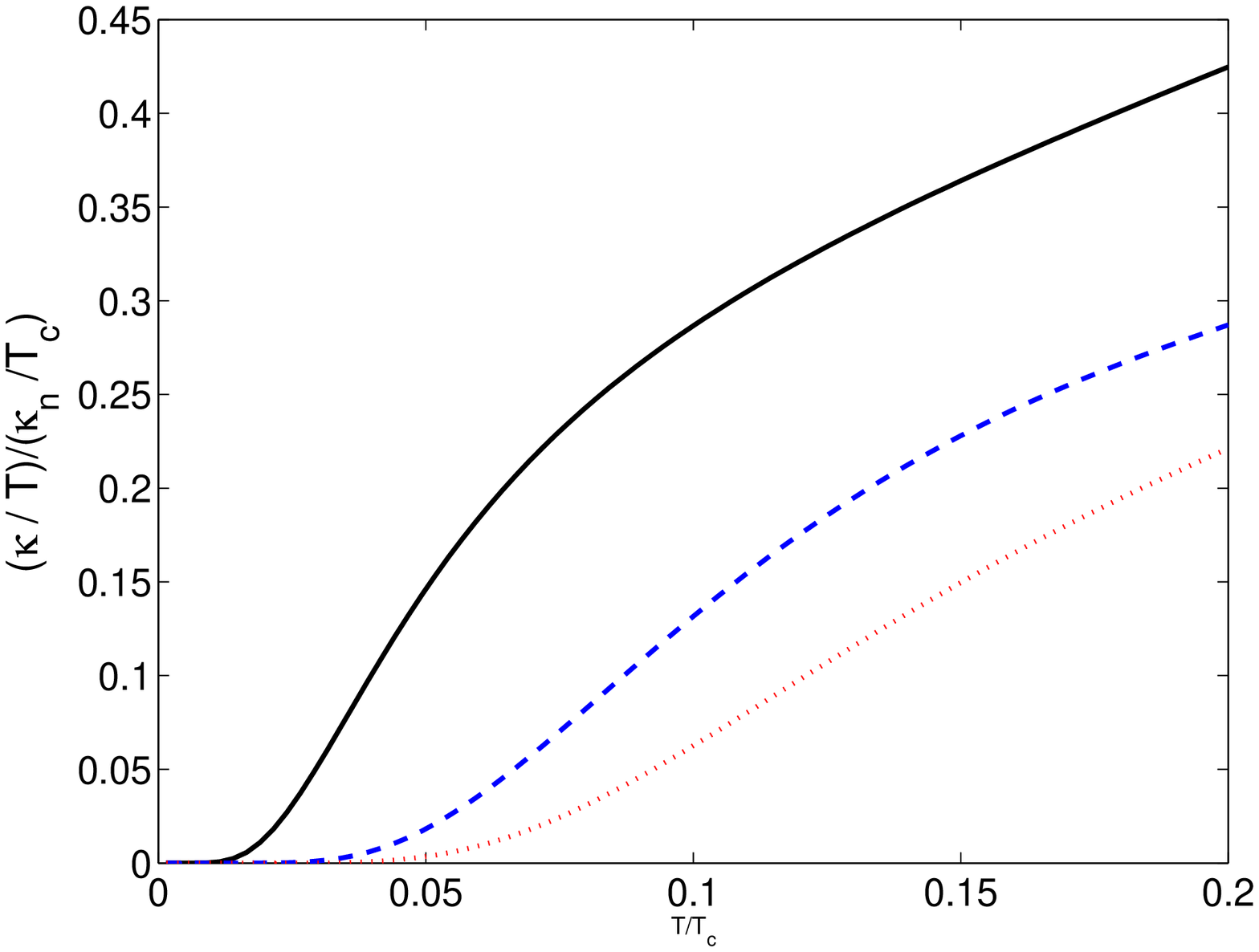}
\caption{(Top)Normalized density of states $N(w)/N_0$ vs.
$\omega/T_{c0}$ for two band anisotropic model with isotropic gap
on sheet 1 and $r=0.9$ (deep gap minima) on sheet 2 (Eq.
(\ref{eq:gps}). Results are shown for various values of intraband
scattering rate $\Gamma/T_{c0}$ and $c=0.07$. Bottom: normalized
thermal conductivity $(\kappa(T) / T) /(\kappa_n / T_c)$ vs.
$T/T_c$.}
\label{fig:intra_k}
\end{center}
\end{figure}

{An alternative
scenario for the absence of a linear term in $\kappa(T)$ in {K-,
Ni- and Co-doped} 122 ferropnictide materials
\cite{Tailleferkappa,Likappa,TanatarCo:2009}, is a highly
anisotropic state on at least one of the Fermi surface sheets with
deep minima but no true nodes.} Here we choose $\Delta_{ani} <
\Delta_{iso}$ {($r=0.9$)} to simulate a situation where  the clean
state is slightly gapped, and varies between a minimum value of
$\Delta_{iso}-\Delta_{ani}$ and $\Delta_{iso}+\Delta_{ani}$. Again
we begin by  including only intraband impurity scattering, for
which the results are shown in Fig. \ref{fig:intra_k}.  In this
case, disorder merely increases the spectral gap due to averaging,
as discussed in Ref. \onlinecite{Mishraetal:2009}, leading to an
increasingly rapid exponential decay.

\begin{figure}
\begin{center}
\includegraphics[width= 0.9\columnwidth]{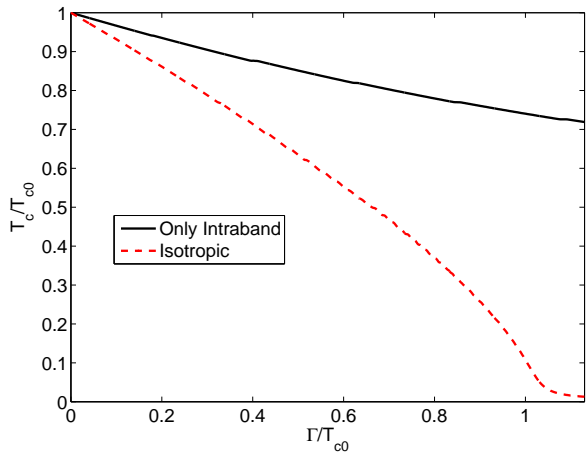}
\caption{Comparison of the effects of intraband and isotropic
scattering on $T_c$.  Solid curve is pure intraband scattering
$T_c/T_{c0}$ vs. $\Gamma/T_{c0}$ for $U_{12}=0$; dashed curve is
same, but for $U_{12}=U_{11}=U_{22}$.}
\label{fig:Tc}
\end{center}
\end{figure}

\begin{figure}
\begin{center}
\includegraphics[width= 0.9\columnwidth]{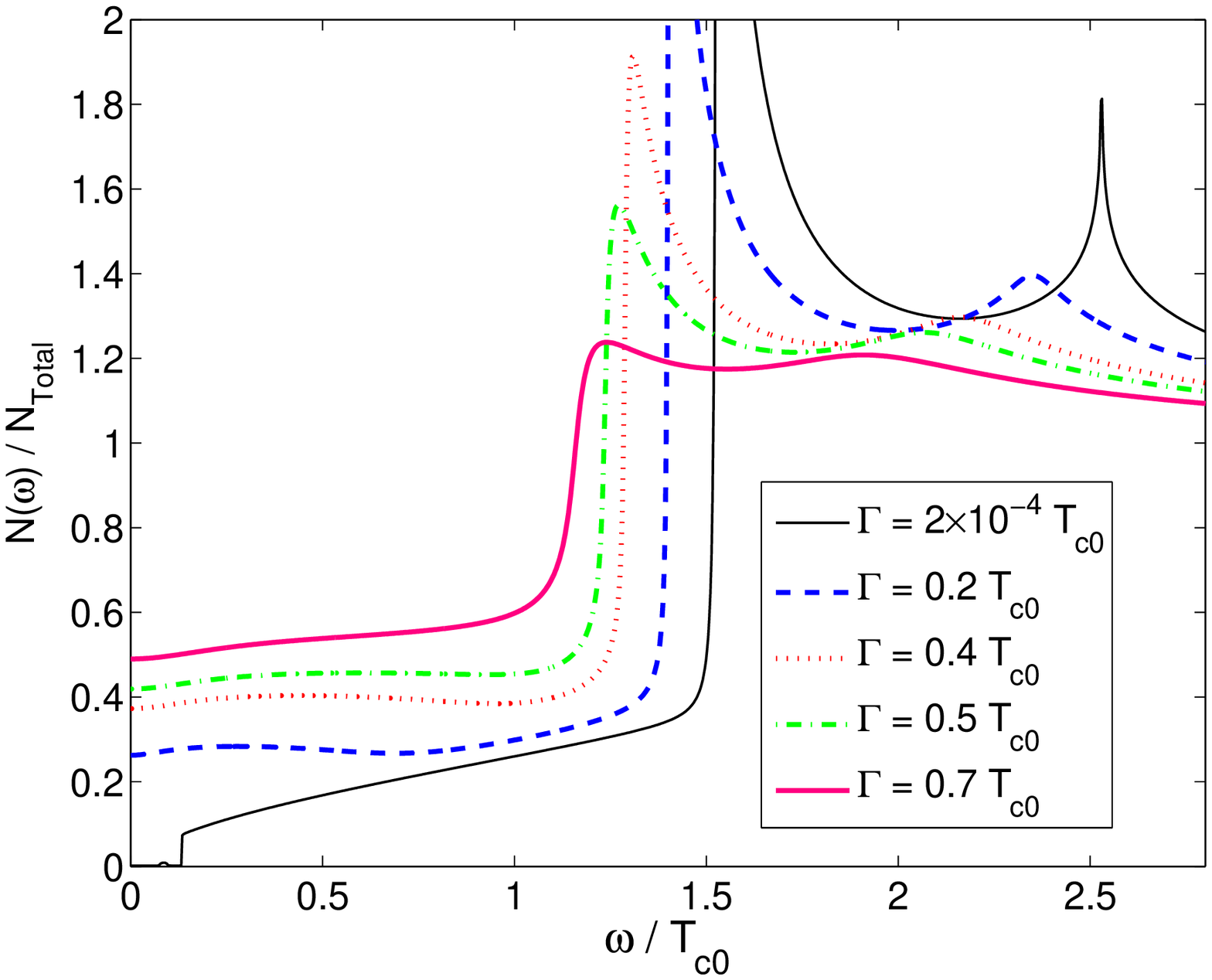}
\includegraphics[width= 0.9\columnwidth]{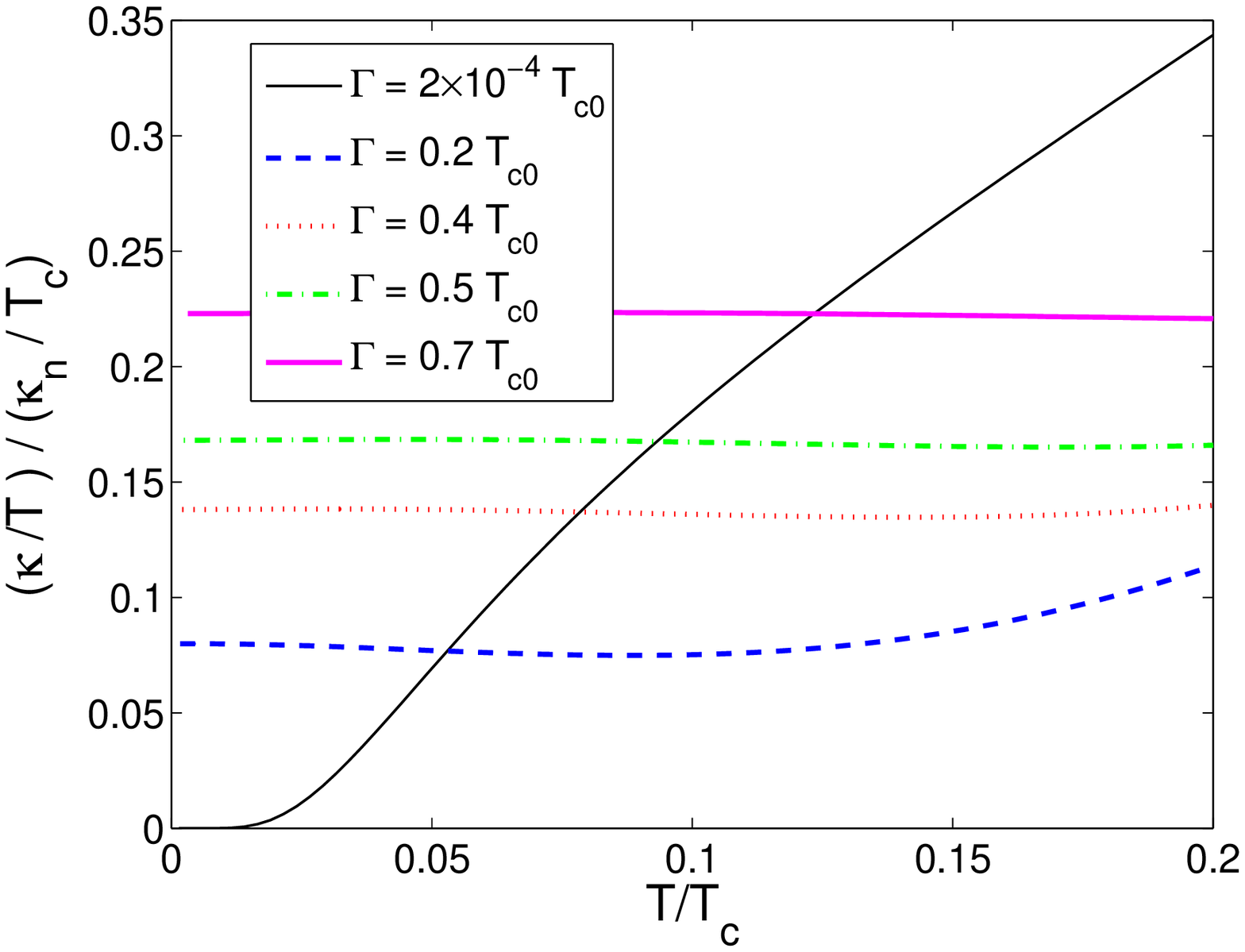}
\caption{Density of states $N(w)/N_0$ (top) and normalized thermal
conductivity $\kappa(T) T_c/(\kappa_n T)$ vs. $T/T_c$ for two band
anisotropic model with isotropic order parameter on sheet 1 and
$r=0.9$ (deep gap minima) on sheet 2 (Eq. (\ref{eq:gps}). Results
are shown for equal intraband and interband scattering
$U_{12}=U_d$.} \label{fig:kappa_nonode}
\end{center}
\end{figure}

\begin{figure}
\begin{center}
\includegraphics[width= 0.8\columnwidth]{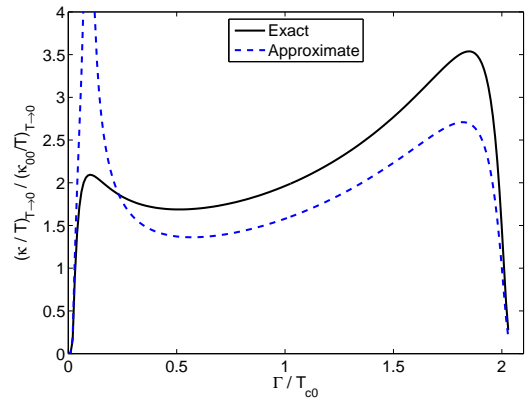}
\caption{$\kappa / T$ in $T\rightarrow0$ limit, for the state with
deep gap minima. $\kappa_{00}$ is thermal conductivity for pure
d-wave state. Band 1 has isotropic order parameter $\Delta=-1.7~T_{c0}$
and band 2 has $r=0.9$ and $\Delta_{iso}=1.5~T_{c0}$ on it.}
 \label{fig:uni2}
\end{center}
\end{figure}

If interband scattering is included, low-energy states appear,
similarly to the $s_\pm$ case considered above, but again we find
that the values of interband scattering strength $U_{12}$ near the
intraband value {$U_{d}$} are necessary to create such states
sufficiently near the Fermi level to create a linear term in
$\kappa$. For simplicity, therefore, we take both inter and intra
band scattering potentials to be equal, i.e. the scattering is
isotropic in momentum space. For intermediate to strong
potentials, states {are then} created near the Fermi level. It is
important to remember that such scattering rapidly suppresses
$T_c$, as illustrated in Fig. \ref{fig:Tc}.
The next figure exhibits the thermal conductivity for this system.
Because the pure system has a small spectral gap, even the
smallest {disorder in this limit} gives rise to an impurity band
close to the Fermi level, {creating unpaired quasiparticles}. For
significant {impurity} concentrations, a strong linear term
appears which also of course violates universality, as shown
explicitly in Fig. \ref{fig:kappa_nonode}. The variation of this
linear term with disorder for isotropic scattering is shown in
Fig. \ref{fig:uni2} and {compared to an approximation where we
made a series expansion around $\pi/2$ for $\Delta_\phi$.  We
find}

%
%
%
%
%
%
\begin{eqnarray}
 \frac{\kappa}{T} &\approx & \frac{N_2 v_{F2}^{2}\pi}{6} \frac{\Gamma_{2}^{~2}}{\Gamma_{~2}^{2} + (\Delta_{iso} +
 \Sigma_{2,1}-\Delta_{ani})^2} \nonumber
 \\ &&\times \frac{1}{\sqrt{\Delta_{ani}~(\Delta_{iso} + \Sigma_{2,1}-\Delta_{ani})}}
 \nonumber
\\   && +  \frac{N_1 v_{F1}^{2} \pi^2}{6}\frac{\Gamma_{1}^{~2}} {(\Gamma_{1}^{~2} + (\Delta +\Sigma_{1,1})^{2} )^{3/2} } \,,  \label{eq:uni2}
 \label{kappa_unitary_minima}\end{eqnarray}
{where $\Gamma_i$ are the normal state scattering rates defined
above, and all the self-energies are evaluated at $\omega=0$.}
Note that there now appears a contribution from the isotropic band
1 because Eq. \ref{kappa_unitary_minima} assumes the isotropic
scattering condition $U_{11}=U_{12}$, which leads to strong
pairbreaking and quasiparticle states near the Fermi level.
{Consequently, the absence of the residual linear term in $\kappa$
is also consistent with the deep minima provided the interband
scattering is not too strong. We now proceed to investigate the
field dependence in the two cases.}

\section{Field dependence}

Thermal conductivity depends on the applied magnetic field since
the density of unpaired electrons depends on the field magnitude.
These electrons carry entropy and hence enhance the heat current.
They also scatter phonons and therefore reduce the lattice
contribution to the thermal transport, so that the two effects
compete. On general grounds, $\kappa(T,H)$ that increases at low
temperatures with applied field can be assumed to contain a
substantial electronic component \cite{YMatsuda:2006}. In some
systems, such as heavy fermion metals, the electron contribution
to the thermal conductivity is dominant, allowing a direct probe
of the heat transport in the superconducting state throughout the
$T$-$H$ plane \cite{RMovshovich:2001}. In other materials, where
the phonon contribution is substantial, the quantity that lends
itself most easily to analysis is the {\em residual} linear term
in the thermal conductivity, $\lim_{T\rightarrow 0}\kappa/T$,
which is purely electronic since the phonon contribution vanishes
in that limit
\cite{LTaillefer:1997,EBoaknin:2003,MSutherland:2003}.

Therefore, for the purposes of comparison with experiment, we
focus on the field dependence of the electronic thermal
conductivity at low temperature. In nodal superconductors, where
the transport is dominated by bulk quasiparticles with momenta
nearly along the nodal directions, two methods have been employed
to describe this dependence. The semiclassical approach is based
on the observation that the energy to break a Cooper pair is
lowered outside of the vortex core since the unpaired electrons do
not participate in the supercurrent flowing around the vortex.
Hence effect of the field can be described by the Doppler shift of
the quasiparticle energy, ${\bm v}_s(\bm r)\cdot \bm k_F$, where
$\bm v_s(\bm r)$ is the supervelocity field determined by the
vortex structure
\cite{GVolovik:1993,CKubert:1998SSC,IVekhter:2001}. This energy
shift is local, and therefore the method is very well suited for
describing the thermodynamic quantities, but requires additional
assumptions to account for correlation functions and transport
properties
\cite{CKubert:1998,MFranz:1999,IVekhter:2000anis,IVekhter:2000}.
It is applicable at low energies and therefore restricted to low
temperatures and fields.

An alternative approach assumes the existence of the vortex
lattice and describes the behavior starting from the moderate to
high field regime. The approximation consists of replacing the
diagonal, in Nambu space, components of the Green's function with
their averages over the unit cell of the vortex lattice, while
keeping the exact spatial dependence of the off-diagonal, Gor'kov
components. It was developed for conventional superconductors by
Brandt, Pesch, and Tewordt \cite{BPT:1967,WPesch:1975}, who showed
that the replacement is valid since the Fourier components of the
Green's function (in reciprocal lattice vectors of the vortex
lattice, $\bm K$), vary as $G_{\bm K}\propto \exp(-\Lambda^2K^2)$,
where $\Lambda^2=\hbar c/eB$ is the magnetic length, which is of
order of the intervortex distance. The method gives excellent
agreement with experimental results on both thermodynamic and
transport properties superconductors near the upper critical field
\cite{AHoughton:1971,PKlimesch:1978}, and remains
semi-quantitatively correct in $s$-wave superconductors down to
fields of less then half of $H_{c2}$ \cite{EHBrandt:1976}. It
fails at the lowest fields, when the unpaired electrons are
localized in the vortex cores, and consequently cannot be
described by the propagators averaged over the (much greater) unit
cell size.  In this low field limit, the method gives and
artificially enhanced behavior of the thermal conductivity as it
treats the localized states as extended, and generically produces
power law increase in $\kappa(H,T)$, while both the expected and
the experimentally observed initial increase in $\kappa(H,T)$ is
exponentially small.

In contrast, the extension of the BPT method to the nodal
superconductors
\cite{IVekhter:1999,HKusunose:2004,AVorontsov:2006,ABVorontsov:2007a,ABVorontsov:2007b}
remains valid down to lowest fields since even in that regime the
transport is dominated by the extended states. It gives results
qualitatively and quantitatively consistent with the Doppler shift
method \cite{TDahm:2002,GRBoyd:2009}, and hence describes the
properties over nearly the entire range of the temperatures and
fields. We employ this method here.

We extend the approach or Ref.\onlinecite{ABVorontsov:2007b} to
the two-band model. The matrix Green's functions of the electron
and hole bands are coupled by the self-consistency equation on the
order parameter and  the $T$-matrix as shown in Eq.(\ref{eq:gp})
and Eqs.(\ref{Sigma10})-(\ref{SigmaD}). We model the vortex
lattice as \cite{AVorontsov:2006,ABVorontsov:2007a}
\begin{equation}
  \Delta_i(\bm R, \phi)=\sum_{k_y} C_{k_y}
  {\Phi}_i(\phi) e^{ik_y y}
     F_0\left( {x-\Lambda^2 k_y \over \Lambda} \right) \, ,
    \label{AVL:s}
\end{equation}
where $F_0$ is the ground state oscillator wave function, the
coefficients $C_{k_y}$ determine the structure of the vortex state
and the amplitude of the order parameter, and $i$ is the band
index. Since the bands are treated as separate, the thermal
conductivity is the sum of the contributions due to the hole and
the electron sheets of the Fermi surface.

\begin{figure}
\begin{center}
\includegraphics[width= \columnwidth]{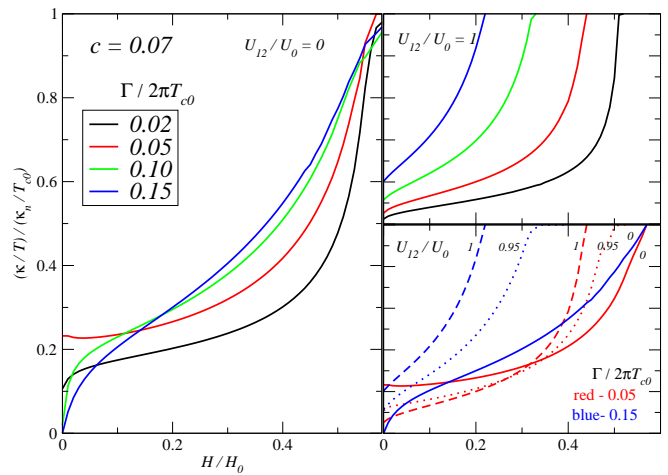}
\caption{(Color online) The field dependence of the  thermal
conductivity at $T=0.02T_{c0}$ for the order parameter with nodes,
$r=1.3$. Left panel: Intraband scattering only for different
scattering rates near the unitarity limit. Right panel top: the
same in the presence of strong interband scattering. Right panel
bottom: evolution of the field dependence with increased interband
scattering.} \label{fig:kHnodal}
\end{center}
\end{figure}

Our approach is most reliable for the states with nodes in the gap
function, when the results can be trusted over essentially the
entire field range. Note that even though the hole band is always
{fully gapped} in our analysis, at low fields the dominant
contribution is from the electron band with gap nodes. The main
feature in the field dependence, as shown in
Fig.~\ref{fig:kHnodal} is the pronounced inflection point at low
fields where a crossover from a rapid rise to a slower increase
occurs. This result bears striking resemblance to the recent
measurements on LaFePO superconductor \cite{Matsudakappa}, which
were interpreted precisely in the framework of the two-band
picture, with one band possessing nodes in the gap.

This rapid increase is related to a significant variation in the
density of states with energy in zero field, shown in
Fig.~\ref{fig:kappa}. The corresponding thermal conductivity for
the parameter values chosen here also exhibits a shoulder as a
function of temperature at low $T$.
This shoulder is not found experimentally in
Ref.~\onlinecite{Matsudakappa}, but we have to keep in mind that,
if the sample contains non-superconducting regions, the residual
linear term may not be related to the superconducting phase, while
the field enhancement at low $T$ is still determined by the
increase in the number of unpaired electrons.

Note also that when the residual linear-$T$ term is near its
maximal value in Fig.~\ref{fig:kappa_T0}, which for parameters
here occurs near $\Gamma/2\pi T_{c0}\simeq 0.05$, the field and
temperature dependence of $\kappa(T,H)$ is very weak, more
reminiscent of that of a fully gapped superconductor. The nodes
however are not lifted up to a higher impurity concentration,
$\Gamma/2\pi T_{c0}\simeq 0.08$.{ The reason for this is clear
from comparisons with Ref.\onlinecite{Mishraetal:2009}: as the
slope of the gap near the node becomes small, the shape of the gap
function deviates from a simple cosine, and becomes very steep
beyond the near-nodal region, so that moderate field essentially
does not excite additional quasiparticles.}

In general, we do need to keep in mind that, as the nodes are
lifted, the applicability of the BPT approximation at low fields
becomes questionable, but the minimal gap in the regime we show
here, $\Gamma/2\pi T_{c0}\leq 0.15$, is small, and therefore the
method remains reliable to very low fields. Quite generally, the
energy scales associated with the effect of magnetic field on the
extended quasiparticles are of order $E_0\sim
\Delta\sqrt{H/H_{c2}}$, so that for the $\Delta_{min}/\Delta\sim
0.1$ we expect the extended states to dominate the response for
$H/H_{c2}\geq 0.01$. Consequently, we trust the approximation even
in the regime when the nodes are lifted.

\begin{figure}
\begin{center}
\includegraphics[width=\columnwidth]{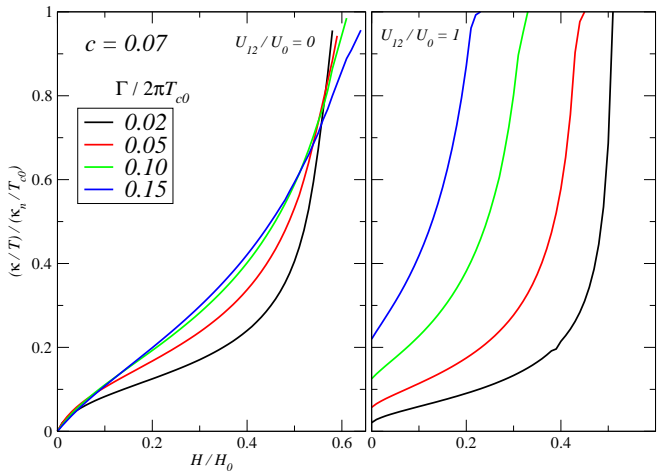}
\caption{(Color online) The field dependence of the  thermal
conductivity at $T=0.02T_{c0}$ for the order parameter with deep minima,
$r=0.9$. Left panel: Intraband scattering only. Right panel: the
strong interband scattering. } \label{fig:kHdeep}
\end{center}
\end{figure}

This argument allows us to extend the treatment to the state with
the deep minima, rather than the true nodes. We consider once
again $r=0.9$ and show the results in Fig.~\ref{fig:kHdeep}. As
can be expected from a probe that is sensitive to the amplitude,
rather than the phase, of the gap function, the overall features
are quite similar to those for a true nodal gap. The low-field
inflection and the rapid rise are not as clearly pronounced,
consistent with the absence of true nodal quasiparticles until the
field is sufficiently high. As pointed out above, for the nodeless
state the inclusion of strong interband scattering leads to a
rapid enhancement in the residual density of states, and the
concomitant increase in the residual linear term in the thermal
conductivity.

\begin{figure}
\begin{center}
\includegraphics[width=\columnwidth]{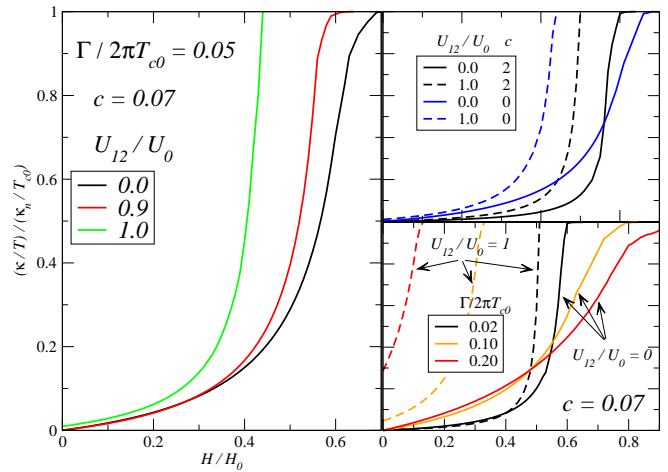}
\caption{(Color online) The field dependence of the  thermal
conductivity for the $s_\pm$ state. Left panel: clean case near
the unitarity limit. Right panel top: influence of the phase shift
of scattering on the low-field behavior. Right panel bottom:
evolution of the field dependence with interband scattering.}
\label{fig:kHs}
\end{center}
\end{figure}

Finally, in Fig.~\ref{fig:kHs} we show the results for the
isotropic {$s_\pm$ state}.  While in this case we do not trust our
approximation at low fields, it is clear that the increase in the
thermal conductivity with the magnetic field is much slower than
for the two cases considered above. To require an even moderately
rapid growth of $\kappa(H)$ at low fields requires a substantial
residual linear term as well as unphysically high interband
scattering, see bottom right panel of Fig.~\ref{fig:kHs}. {This
result strongly suggests that the isotropic $s_\pm$ state is
incompatible with the results of Ref.\cite{Tailleferkappa} on the
122 series of materials, just as the results of
Ref.~\cite{Matsudakappa} exclude this order parameter structure in
the LaFePO system.}

\section{Conclusions}

\label{sec:concl}

We have argued that thermal conductivity is the ideal probe to
resolve current apparent discrepancies between various
thermodynamic and spectroscopic measurements on the Fe-based
superconductors.  In particular, since several of these
experiments indicate the existence of low-lying quasiparticle
states in certain materials, it is important to settle whether or
not these excitations extend all the way down to the Fermi energy,
or whether there is a true spectral gap in the system. Thermal
conductivity, a bulk probe, is currently measurable to lower
temperatures than other probes, so it may be able to settle this
dispute and also distinguish between two popular scenarios. The
two most likely order parameters for these systems appear at
present to be the isotropic $s_\pm$ state proposed by Mazin et
al.\cite{ref:Mazin_exts}, and highly anisotropic $A_{1g}$ states,
with nodes or deep gap minima, found in spin fluctuation
calculations. The former state can be consistent with the reports
of low-lying excitations only if pairbreaking disorder induces an
impurity band, while the latter are difficult to reconcile with
ARPES experiments indicating a large spectral gap in some
materials.

We have therefore calculated the thermal conductivity of a
superconductor with $A_{1g}$ symmetry order parameter in a model
2-band system, and considered the effects of intra- and interband
disorder.  In zero field, we have shown that the linear $T$ term
which dominates $\kappa(T)$ at low temperatures has a coefficient
which is nonuniversal (unlike the well-known $d$-wave case) and
depends non-monotoically on disorder. Details depend on the
precise order parameter structure of the model pure system, and on
location of the impurity band in the DOS. 
{The linear-$T$ term in zero field found in the LaFePO
material~\cite{Matsudakappa}, is consistent in principle with the
linear-$T$ penetration depth observed in the same
system\cite{ref:Fletcher} and suggestive of nodes in the
superconducting order parameter. This term is rather large as a
fraction of the normal state thermal conductivity. While within
the present theoretical approach realistic evaluation of the
normal state $\kappa$ is difficult, as we have neglected both
phonons, which would enhance this contribution, and inelastic
electronic scattering, which would suppress it, it seems likely
that it would be difficult to account for its size within the
current framework, and it is possible, as
Ref.~\onlinecite{Matsudakappa} mentions, that it is of extrinsic
origin.} In the 122 systems, the extremely small or zero
linear-$T$ term\cite{Tailleferkappa,Likappa} suggests a true
spectral gap, consistent either with an isotropic $s_{\pm}$ state
or a gap with deep minima.

An examination of the field dependence of the low temperature
thermal conductivity within the BPT approach has enabled us to
draw further conclusions.  The size of the initial field
dependence seen in experiment rules out a clean $s_\pm$ state as a
possible candidate for the 122 materials.  However, as pointed out
in the context of other experiments, pairbreaking scattering can
induce a low-lying impurity band in such a state, and produce
responses similar to highly anisotropic states.  We have analyzed
this situation and found that the amount of pairbreaking
(intraband) scattering required to reproduce the observed field
dependence is {large. In most situations this is unphysical, both
in the sense that the ratio of interband to intraband scattering
must be tuned to a special value (which seems unreasonable in the
context of screened Coulomb scattering), and because a very large
concomitant $T_c$ suppression would be produced. As mentioned
above, an argument for a sizeable interband scattering component
may be made for Co-doped systems, but the general argument against
fine-tuning to a special value still holds, and the experimental
agreement between the behavior of the thermal conductivity on
systems with different dopants indicates the generic features of
the material. The evolution of the thermal conductivity with Co
doping \cite{TanatarCo:2009}, however, may be at least in part due
to the strong interband scattering component.} We therefore
conclude that the most likely candidate for the order parameter in
the 122 materials is a highly anisotropic $A_{1g}$ state with deep
gap minima, probably on the electron (``$\beta$") sheets.  How
this conclusion can be reconciled with ARPES experiments is not
clear at this writing. {We emphasize that controlled disorder not
associated with doping, such as electronic irradiation, would
provide the best test of the predictions of our theory.}

\acknowledgements  The authors are grateful for  useful
communications with L. Taillefer and Y. Matsuda. PJH is grateful for 
the hospitality of the Kavli Institute for Theoretical Physics during
the preparation of this manuscript.  Research was  partially supported 
by DOE DE-FG02-05ER46236 (PJH) and DOE  DE-FG02-08ER46492 (IV).

\appendix

\label{sec:app}

\section{Basic formalism} 
\begin{figure}
\begin{center}
\includegraphics[width= 0.8\columnwidth]{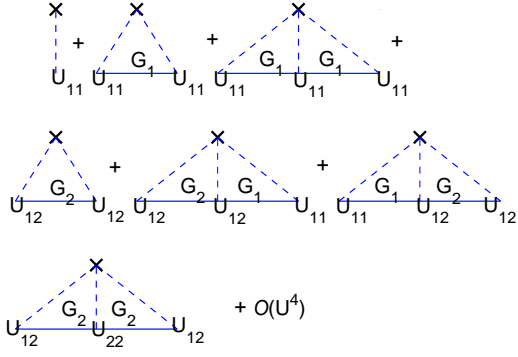}
\caption{These are the impurity averaged diagrams, which
contribute to the self energy of the first band Green's function.
Here the interband contribution comes through processes, which
involve even number of interband scatterings. The diagrams also
takes into account the order  of inter and intraband scatterings.
$U_{ij}$ is the impurity potential strength, where$i,j$ are the
band indexes. $i=j$ denotes the intraband and $i\neq j$ denotes
the interband potential strength. $G_{i}$ is the bare Green's
function.
}
\label{fig:tmat}
\end{center}
\end{figure}

For low impurity concentrations, one
can ignore the processes which involve scattering from multiple
impurity sites. Within this single site approximation, we sum all
possible scattering events from a single site to calculate the
disorder-averaged $t$-matrix, which is then related to the
one-electron self energy as,
\begin{equation}
\widehat\Sigma (\k, \omega)= n_{imp} \widehat T_{\k,\k}(\omega)\,,
\end{equation}
where $n_{imp}$ is the impurity concentration, and, for isotropic
scatterers, $\widehat T_{\k,\k}(\omega)=\widehat T(\omega)$, and
hence $\widehat\Sigma (\omega)$, has no momentum dependence. For a
multiband superconductor, scattering from impurities can occur
within a given band with potential $U_{ii}$, or between the bands
with potential $U_{ij}$, $i\ne j$. Fig. \ref{fig:tmat} shows the
impurity averaged diagrams for the self energy $\Sigma$ which
occur up to third order for a two-band system. Any process which
involves odd number of interband scatterings does not contribute
to the self energy, because the Green's function and self-energy
in the translationally invariant disorder-averaged system are
diagonal in band index.


The sum of all the diagrams involving a single impurity site can
be expressed compactly as

\begin{equation}
\hat T_{i}=\frac{1}{1-\hat U^{eff}_{i}\langle \hat G_i
\rangle_{k}} \hat U^{eff}_{i} \label{eq:Tmt}
\end{equation}
where the effective impurity potential for $i^{th}$ band  is,
\begin{equation}
\hat U_{i}^{eff}=\hat U_{ii}+\hat U_{ij} \langle \hat G_{j}
\rangle_{k} \hat U_{ji}
\end{equation}
In a Nambu basis,
\begin{eqnarray}
\hat U_{ij} &=& U_{ij}~\tau_{3} \\
\langle \hat G_{i} \rangle_{k} &=& g_{i,0}~\tau_{0} +
g_{i,1}~\tau_{1} \label{eq:notation}
\end{eqnarray}
Here {\bf $g_{i,\alpha}$} are the Nambu components of the
momentum-integrated Green's function. The first subscript
$i(=1,2)$ in $g_{i,\alpha}$ stands for the band and the second
subscript $\alpha(=0,1,2,3)$ represents the Nambu channel.

The effective potential for the first band may be written

\begin{equation}
\hat U_{11}^{eff}=U_{11} \tau_3 + U_{12}^2 \tau_{3} \left[\hat{G}_{2}
\frac{1}{1-\hat{U}_{22} \hat{G}_2}\right] \tau_{3} \label{eq:nw_ver}
\end{equation}
In the above equations, $U_{11},U_{22}$ are the intraband impurity
potential strengths in band $1$ and $2$ respectively, while
$U_{12}$ is the interband impurity potential strength. The
$t$-matrix for, e.g. the first band may now be written as

\begin{equation}
\hat T_1=\frac{1}{1- \hat U_{11}^{eff}  \hat G_1} (\hat
U_{11}^{eff}) \label{eq:nw_ver_1}
\end{equation}

After some Pauli matrix algebra, we find that  the self energy
components in band and Nambu channels may be written as

\begin{eqnarray}
\Sigma_{1,0}&=& \frac{n_{imp}}{{\mathcal D}} \left[ U_{11}^2
g_{1,0}+U_{12}^{2} g_{2,0} \right. \label{Sigma10} \\ \nonumber
  &-& \left. g_{1,0}(U_{12}^{2}-U_{11}~U_{22})^{2}~(g_{2,0}^{2}-g_{2,1}^{2}) \right] \\ [2 ex]
\Sigma_{1,1}&=& -\frac{n_{imp}}{{\mathcal D}} \left[ U_{11}^2
g_{1,1}+U_{12}^{2} g_{2,1} \right. \label{Sigma11} \\ \nonumber
  &-& \left. g_{1,0}(U_{12}^{2}-U_{11}~U_{22})^{2}~(g_{2,0}^{2}-g_{2,1}^{2}) \right] \\ [2 ex]
 \Sigma_{2,\alpha}&=& \Sigma_{1,2\rightarrow
 2,1,\alpha}\label{Sigma2a}\\&~&\nonumber\\
 {\mathcal D} &=& 1-2~U_{12}^2 (g_{1,0}~g_{2,0}-g_{1,1}~g_{2,1}) \label{SigmaD}\\ \nonumber
             &+& (U_{12}^{2}-U_{11}U_{22})^2~(g_{1,0}^{2}-g_{1,1}^{2})~(g_{2,0}^{2}-g_{2,1}^{2})  \\ \nonumber
&-&U_{22}^2~(g_{2,0}^{2}-g_{2,1}^{2})-U_{11}^2~(g_{1,0}^{2}-g_{1,1}^{2})
 \label{eq:tmat_simp}
\end{eqnarray}
Here again the first subscript in $\Sigma_{i,\alpha}$ represents
the band index and the second subscript $\alpha$ denotes the Nambu
channel.

\section{Special Cases}

\subsection{Born limit of two band case} In the Born limit, we
will keep the terms up to second order in $``U_{ij}"$, so the
denominator becomes $1$ and we get,

\begin{eqnarray}
\Sigma_{1,0}&=& n_{imp}~(U_{11}^{2}~g_{1,0}+U_{12}^{2}~g_{2,0}) \\
\Sigma_{1,1}&=& -n_{imp}~(U_{11}^{2}~g_{1,1}+U_{12}^{2}~g_{2,2}) \\
\Sigma_{2,0}&=& n_{imp}~(U_{22}^{2}~g_{2,0}+U_{12}^{2}~g_{1,0}) \\
\Sigma_{2,1}&=& -n_{imp}~(U_{22}^{2}~g_{2,2}+U_{12}^{2}~g_{1,2})
\label{eq:selfn_2_born}
\end{eqnarray}

\subsection{Strong potential  limit} Due to the presence of an
additional band, a new parameter comes into play in the unitary
limit, the ratio of the interband scattering to the intraband
scattering. This leads to three distinct cases in the large
potential limit. \vskip .3cm
\paragraph{Strong potential limit I: $U_{11}=U_{22} > U_{12}$} In this
case, intraband scattering dominates, and in the limit the self
energies reduces to

\begin{eqnarray}
\Sigma_{1,0}&=& -n_{imp}~\frac{g_{1,0}}{(g_{1,0}^2-g_{1,1}^2)} \\
\Sigma_{1,1}&=& n_{imp}~\frac{g_{1,1}}{(g_{1,0}^2-g_{1,1}^2)}  \\
\Sigma_{2,0}&=& -n_{imp}~\frac{g_{2,0}}{(g_{2,0}^2-g_{2,1}^2)} \\
\Sigma_{2,1}&=& n_{imp}~\frac{g_{2,1}}{(g_{2,0}^2-g_{2,1}^2)}
\label{eq:selfn_unitary_1}
\end{eqnarray}
It is clear that in this limit, states within a given band are
broadened only by interband scattering processes. \vskip .3cm

\paragraph{Strong potential limit II: $U_{11}=U_{22} = U_{12}$.}  In this very special case,
 for strong potentials the self energies
become

\begin{eqnarray}
\Sigma_{1/2,0}&=&  -n_{imp}~\frac{g_{1,0}+g_{2,0}}{(g_{1,0}+g_{2,0})^2-(g_{1,1}+g_{2,1})^2} \\
\Sigma_{1/2,1}&=&  n_{imp}~\frac{g_{1,1}+g_{2,1}}{(g_{1,0}+g_{2,0})^2-(g_{1,1}+g_{2,1})^2} \\
\label{eq:selfn_unitary_2}
\end{eqnarray}
In this case, both the bands have identical self energies.  As
discussed in the text, it corresponds to the presence of bound
states at low energies in the $s_\pm$ state, so we devote some
attention to it.


\begin{thebibliography}{99}
\bibitem{ref:footnote1} We have averaged the $xx$ and $yy$
components to simulate the ferropnictide with order parameters on
$\beta_1$ and $\beta_2$ sheets rotated by $\pi/2$ in local
coordinates with respect to one another.
\end{thebibliography}
\end{document}